\documentclass{article}
\usepackage{spconf,amsmath,epsfig}
\usepackage{spconf,amsmath,graphicx}
\usepackage{subcaption}

\usepackage{booktabs,float,subfig}
\usepackage{epstopdf,booktabs,multicol,multirow,enumitem,footnote,type1cm,amssymb,calc}
\usepackage{overpic}
\usepackage{color,xcolor}
\let\OLDthebibliography\thebibliography
\renewcommand\thebibliography[1]{
  \OLDthebibliography{#1}
  \setlength{\parskip}{0pt}
  \setlength{\itemsep}{0pt plus 0.3ex}
}

\begin{document}\sloppy

\def\x{{\mathbf x}}
\def\L{{\cal L}}

\title{HydraFormer: One Encoder For All Subsampling Rates}
%
\name{Yaoxun Xu$^{1,\dagger}$, Xingchen Song$^{1,3,\dagger}$\thanks{$\dagger$ Equal contribution.  * Corresponding authors.}, Zhiyong Wu$^{1,2,4,*}$, Di Wu$^{3}$, Zhendong Peng$^{3}$, Binbin Zhang$^{3}$}
\address{
  $^1$Shenzhen International Graduate School, Tsinghua University, Shenzhen, China\\
  $^2$Peng Cheng Lab, Shenzhen, China \quad
  $^3$Horizon Inc, Beijing, China \\
  $^4$The Chinese University of Hong Kong, Hong Kong SAR, China}

\maketitle

\begin{abstract}
In automatic speech recognition, subsampling is essential for tackling diverse scenarios. However, the inadequacy of a single subsampling rate to address various real-world situations often necessitates training and deploying multiple models, consequently increasing associated costs. To address this issue, we propose HydraFormer, comprising HydraSub, a Conformer-based encoder, and a BiTransformer-based decoder. HydraSub encompasses multiple branches, each representing a distinct subsampling rate, allowing for the flexible selection of any branch during inference based on the specific use case.
HydraFormer can efficiently manage different subsampling rates, significantly reducing training and deployment expenses. 
Experiments on AISHELL-1 and LibriSpeech datasets reveal that HydraFormer effectively adapts to various subsampling rates and languages while maintaining high recognition performance.
Additionally, HydraFormer showcases exceptional stability, sustaining consistent performance under various initialization conditions, and exhibits robust transferability by learning from pretrained single subsampling rate automatic speech recognition models\footnote{Model code and scripts: https://github.com/HydraFormer/hydraformer}.
\end{abstract}
\vspace{-1mm}
\section{Introduction}
\vspace{-2mm}
\label{sec:intro}

Automatic speech recognition (ASR) \cite{asr0,asr1}, as a crucial branch in the field of artificial intelligence, plays a pivotal role in various application scenarios.
With the continuous advancement of technology, ASR models face the challenges of reducing computational complexity and adapting to diverse application scenarios while improving recognition accuracy. To address these issues, researchers have introduced subsampling techniques, aiming to maintain essential information while reducing data volume and enhancing the model's operational efficiency.

Current ASR research typically uses fixed subsampling rates, yielding satisfactory performance in specific scenarios.  For instance, \cite{subsampling4} and \cite{subsampling4_1} opt for a subsampling rate of 4, while \cite{subsampling3} and \cite{subsamplingrate2} choose rates of 3 and 2, respectively.


\begin{figure}[h]

\begin{subfigure}{.24\textwidth}
  \centering
  \includegraphics[width=1.0\linewidth]{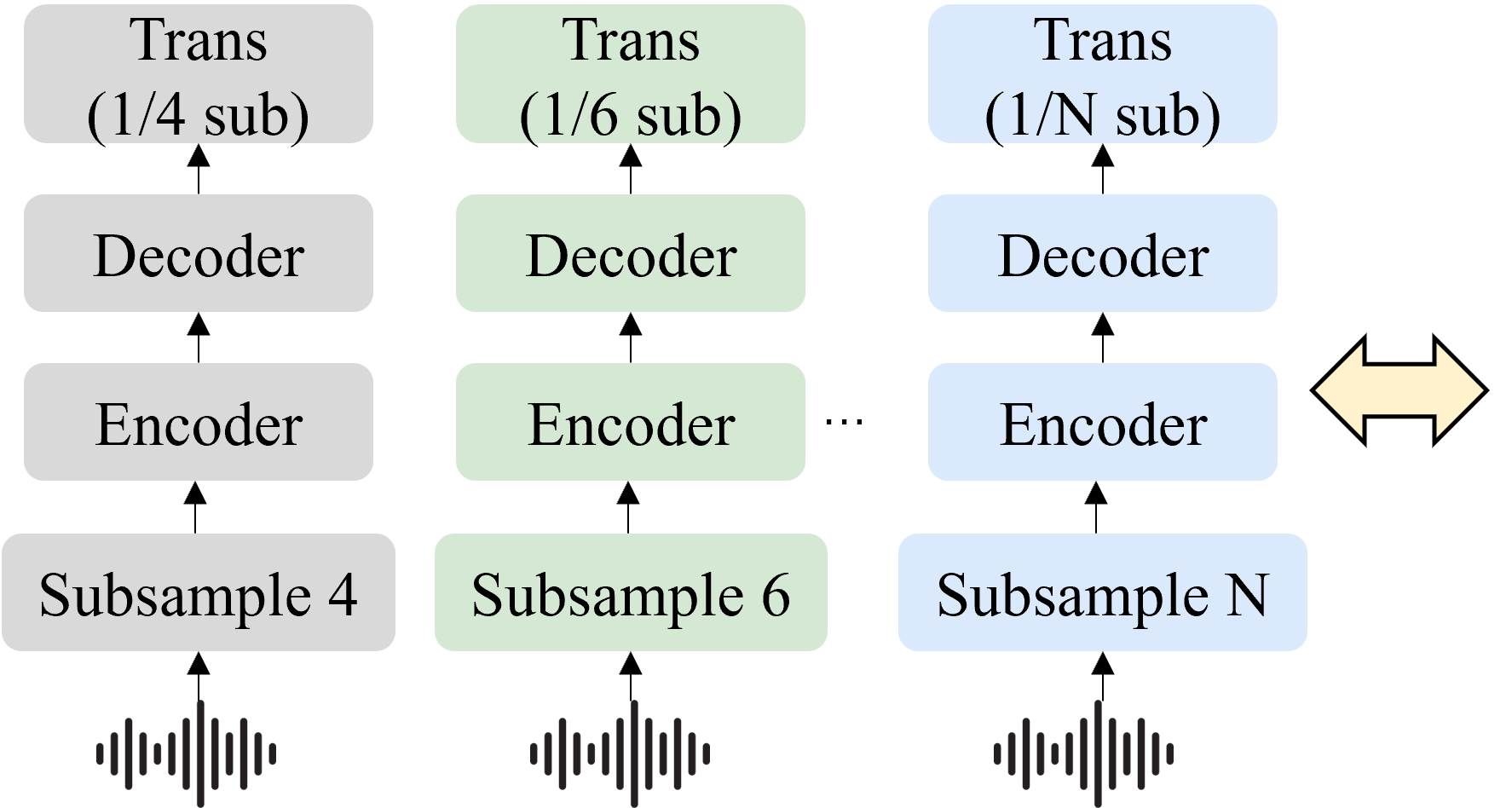}
  \caption{Single subsampling rate models}
  \label{fig:sub1}
\end{subfigure}%
\begin{subfigure}{.24\textwidth}
  \centering
  \includegraphics[width=1.0\linewidth]{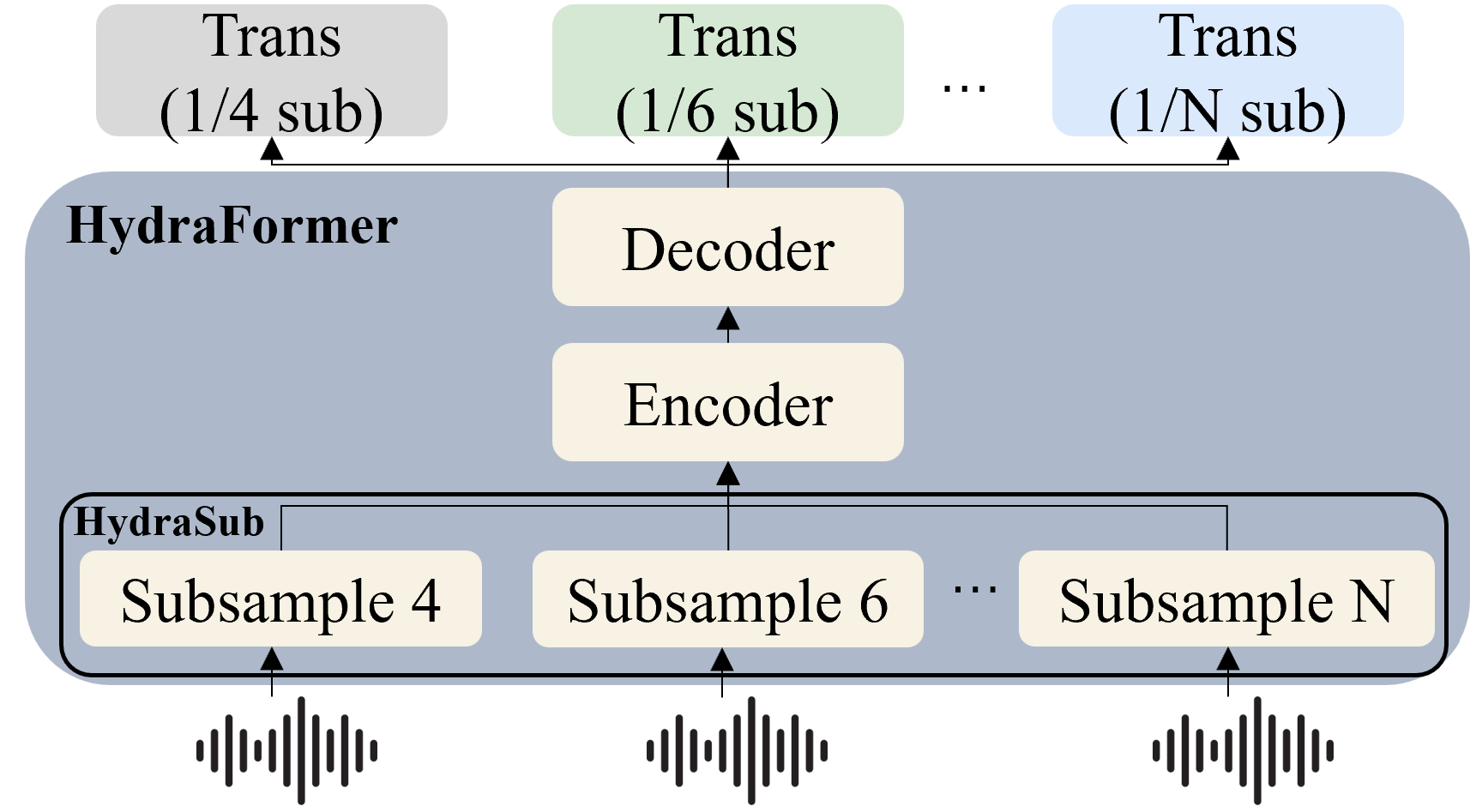}
  \caption{HydraFormer }
  \label{fig:sub2}
\end{subfigure}
\caption{Comparison of HydraFormer and single subsampling rate ASR models}
\vspace{-7mm}
\label{fig:test}
\end{figure}

In real-world applications, certain scenarios prioritize real-time factor (RTF) over word error rate (WER). For example, edge-cloud hybrid recognition systems \cite{edgecloud} emphasize speed on edge-side speech recognition and accuracy on cloud-side speech recognition. Using the same subsampling rate model for both sides may degrade performance, often necessitating two different models. Additionally, situations like slow speech recognition \cite{slow} benefit from aggressive subsampling strategies, improving RTF and maintaining WER with larger subsampling rates. Moreover, variations in languages, accents, and speakers can yield suboptimal performance for fixed-rate models in some circumstances.

These scenarios emphasize that a single subsampling rate may not satisfy diverse real-world application needs, necessitating multiple models for different situations. This approach not only increases training and deployment costs but also adds to model management and maintenance complexity. Furthermore, updating or iterating models requires simultaneous updates across all models, leading to substantial expenses. Consequently, developing a unified model that adapts to the subsampling needs of different scenarios while maintaining high recognition performance has become a pressing issue.
\begin{figure*}[!htbp]
  \centering
  \includegraphics[width=.9\textwidth]{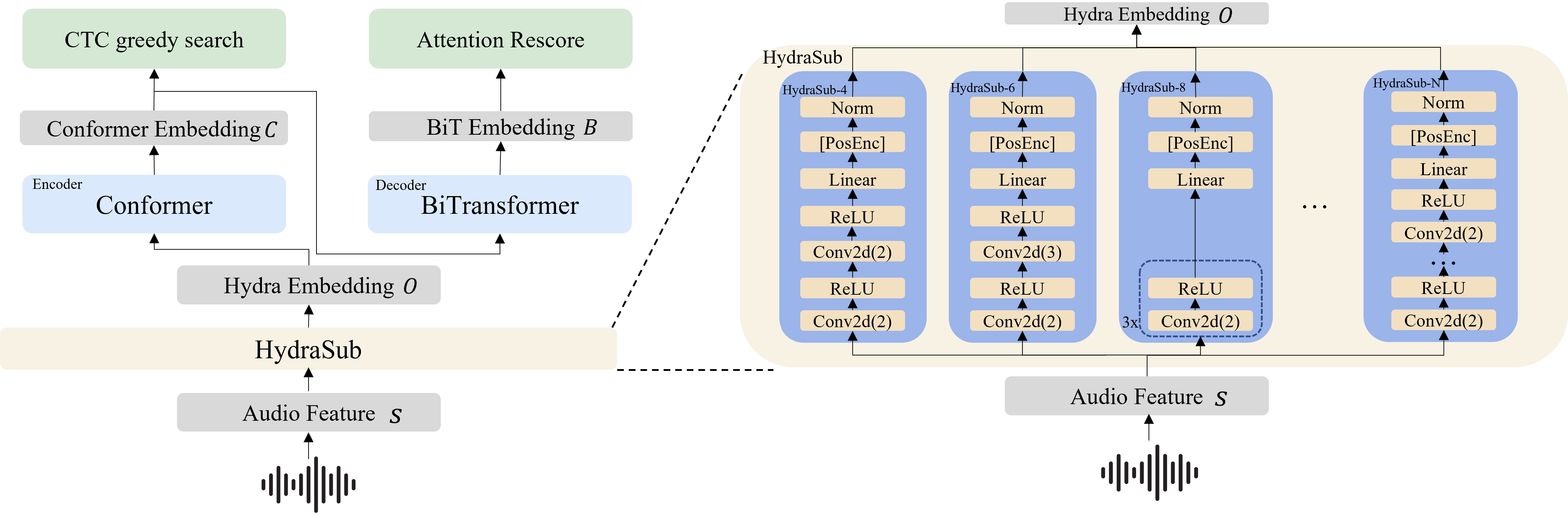}
  
  \caption{Architecture of HydraFormer, where Conv2d (2) denotes convolution layer with stride 2, [PosEnc] denotes optional positional encoding}
  \vspace{-5mm}
  \label{fig:my_label}
\end{figure*}

\cite{zhang2019trainable} proposes a method for achieving speech recognition at different downsampling rates through dynamic frame dropping. However, due to the frame dropping, it inevitably loses a significant amount of information. \cite{han2021multistream} suggests using different subsamples with different encoders, which can achieve various downsampling rates. However, this method requires designing a dedicated encoder for each downsampling rate, which greatly increases the model's parameter size and subsequently the training and deployment costs.  \cite{kim2022squeezeformer} and \cite{yao2023zipformer} adopt a U-Net-like architecture incorporating multiple sampling rates. However, certain limitations are evident in their approaches. Firstly, the U-Net-inspired structure entails a progressive subsampling process, which continues until the rate is reduced to one-eighth of the original. Each subsampling portion within this process is linked to an encoding module that houses a significant number of parameters which results in substantial computational overhead and inefficiency. Moreover, they lack the flexibility to dynamically adjust the subsampling rate during inference, limiting adaptability in diverse scenarios.

To address the problem, we propose HydraFormer, a versatile model capable of adapting to different subsampling rates to meet the demands of various application scenarios, as illustrated in Fig. 1. The key idea behind HydraFormer is to enable a single model to handle $N$ different subsampling rates, effectively replacing $N$ individual models that use single subsampling rates. This approach significantly reduces the training and deployment costs to $\boldsymbol{1/N}$ of the original expense while maintaining comparable performance levels. The main contributions of this paper are as follows:

\begin{itemize}
[itemsep=1pt,topsep=0pt,parsep=0pt]
    \item HydraFormer uses a shared encoder for different subsampling rates, reducing training and deployment costs while maintaining comparable recognition performance at various rates compared to single-subsampling-rate ASR models.
    \item Omitting positional encoding in HydraSub enhances performance, as various subsampling rates with HydraSub may disrupt the encoder's temporal perception.
    \item HydraFormer maintains consistent performance across various model initialization strategies, demonstrating its stability, and it can be easily fine-tuned from pretrained single subsampling rate ASR models.
\end{itemize}

\vspace{-1mm}
\section{Methodology}
\vspace{-2mm}
Fig. 2 shows the primary architecture of HydraFormer, which mainly consists of three components: HydraSub, Conformer-based \cite{gulati2020conformer} encoder, and BiTransformer-based \cite{vaswani2017attention} decoder. We choose the Conformer as the encoder due to its powerful global and local feature extraction capabilities and its high computational efficiency. 
Furthermore, we employ the BiTransformer as the decoder to enhance the receptive field, allowing the model to access more contextual information and subsequently improve its decoding ability.

\vspace{-3mm}
\subsection{HydraSub}
\vspace{-1mm}
As shown in Fig. 2,
HydraSub consists of multiple branches, HydraSub-4, HydraSub-6, HydraSub-8, ..., HydraSub-N, each representing a subsampling factor $n \in \{4, 6, 8, \dots, N\}$. Each branch contains convolutional and ReLU layers to perform subsampling on the time dimension. We represent the input audio features as $\mathbf{S} \in \mathbb{R}^{B \times T \times I}$, where $B$ is the batch size, $T$ is the number of time steps, and $I$ is the input dimension. The output features Hydra Embedding are represented as $\mathbf{O} \in \mathbb{R}^{B \times T' \times D}$, where $D$ is the output dimension and $T' = \lfloor T / n \rfloor$. 

Taking HydraSub-4 ($n=4$) as an example, the subsampling process can be described with the following equations:
\vspace{-1mm}
\begin{equation}
\begin{aligned}
    \mathbf{X}_4 &= \mathbf{S} \oplus (1, 1, I)  \\
    \mathbf{Y}_{4}^{1} &= \text{ReLU}(\text{Conv2d}(\mathbf{X}_4, D, (3, 3), (2, 2)))  \\
    \mathbf{Y}_{4}^{2} &= \text{ReLU}(\text{Conv2d}(\mathbf{Y}_{4}^{1}, D, (3, 3), (2, 2)))  \\
    \mathbf{Z}_4 &= \text{Linear}(\mathbf{Y}_{4}^{2}, D)\\
    \mathbf{O}_4^{'} &= \left\{
        \begin{aligned}
            &\text{PosEnc}(\mathbf{Z}_4) , && \text{if using positional encoding}\\
            &\mathbf{Z}_4, && \text{otherwise}
        \end{aligned}
    \right.\\
    \mathbf{O}_4 &= \text{LayerNorm}(\mathbf{O}_4^{'})
\end{aligned}
\label{eq:hydra_sub4}
\end{equation}
Here, $\oplus$ denotes the unsqueeze operation, $\text{Conv2d}$ represents a 2D convolutional layer, $\text{ReLU}$ is the ReLU activation function, $\text{Linear}$ is a fully connected layer, $\text{[PosEnc]}$ is the optional positional encoding layer (the presence or absence of PosEnc has a significant impact on experimental results; please refer to Section 3.2 for details), and $\text{LayerNorm}$ is the normalization layer. The LayerNorm is added to improve the training stability by normalizing 
the output features $\mathbf{O}^{'}$.

HydraSub-6 and HydraSub-8 exhibit a similar overall structure to HydraSub-4. However, in their convolutional layers, HydraSub-6 employs convolutional strides of 2 and 3, while HydraSub-8 utilizes 
stride of 2 across all three convolutional layers. For other branches, HydraSub-$n$, the structure can be adapted 
based on the specific subsampling factor.

\vspace{-3mm}
\subsection{Dynamic subsample}
\vspace{-1mm}
To enable HydraFormer to adapt to various subsampling rates, we employ an innovative training strategy. In each training batch, we randomly select one of the HydraSub branches, denoted as HydraSub-$n$, where $n \in \{4, 6, 8, \dots, N\}$. To ensure consistent performance across all branches, we implement a balanced training process where each branch is randomly selected through a uniform distribution, guaranteeing that every branch receives equal training opportunities. We feed the output of the selected HydraSub-$n$ branch, $\mathbf{O}_{n}$, into the Conformer-based encoder and obtain Conformer Embedding $\mathbf{C}_n$. Afterward, we pass $\mathbf{C}_n$ through the BiTransformer-based decoder, resulting in BiT Embedding $\mathbf{B}_n$. During the training process of this batch, we only perform forward and backward propagation through the selected HydraSub-$n$ branch, not the other branches.
We only use $\mathbf{C}_n$, and the ground truth transcription encoded into characters or BPE tokens, represented as $\text{T}$, to compute the CTC loss, denoted by $\mathcal{L}_{\text{CTC}}$. Additionally, we use $\mathbf{B}_n$ and $\text{T}$ to compute the KL divergence loss, denoted by $\mathcal{L}_{\text{KL}}$. We employ a weighted sum of these two losses as the overall training loss, represented as:
\begin{equation}
\begin{aligned}
    \mathcal{L}_{\text{total}} = \alpha \mathcal{L}_{\text{CTC}}(\mathbf{C}_n, \text{T}) + (1 - \alpha) \mathcal{L}_{\text{KL}}(\mathbf{B}_n, \text{T})
\end{aligned}
\label{eq:total_loss_i}
\end{equation}
where $\alpha \in [0, 1]$ is the weighting factor. $\mathcal{L}_{\text{total}}$ is used to update HydraSub-$n$, Conformer-based encoder, and BiTransformer-based decoder via the backpropagation algorithm. During this process, the parameters of the other unselected HydraSub branches remain unchanged. We continuously repeat these steps throughout the training process until the model converges.

This training strategy equips HydraFormer with a comprehensive understanding of various subsampling rates' characteristics and the ability to decode at different rates. Owing to its robust generalization and adaptability, HydraFormer can seamlessly select and decode at different subsampling rates during the inference process. In the inference stage, when choosing a specific subsampling rate, we process the audio exclusively using the corresponding HydraSub branch, then forward it to the encoder for additional processing, and ultimately decode it with the decoder.
\setcounter{table}{1}

\vspace{-1mm}
\section{Experiments}
\label{sec:exp}

\vspace{-2mm}
\subsection{Datasets and experimental setup}
\vspace{-1mm}

To thoroughly assess the proposed HydraFormer, we perform experiments on both Chinese and English datasets, specifically testing HydraFormer on the open-source Chinese Mandarin speech corpus AISHELL-1 \cite{aishell} and the English speech corpus LibriSpeech \cite{librispeech}.

We employ the end-to-end speech recognition toolkit WeNet \cite{zhang2022wenet} for experiments. To facilitate a fair comparison, we adopt identical training and testing settings as the open-source AISHELL-1 and LibriSpeech recipes for each experiment, including loss weights (e.g., $\alpha$), model size, model regularization (e.g., weight decay), optimizer, learning rate schedule, and data augmentation.

HydraFormer can be composed of multiple HydraSub branches. In this study, we aim to provide a more focused analysis and comparison by selecting subsampling factors that are frequently used in practice, specifically 4, 6, and 8.

To validate the effectiveness of HydraFormer, we compare it with current mainstream single subsampling rate speech recognition approaches. 
Specifically, we establish three distinct baseline models, each employing a single fixed subsampling rate (4, 6, or 8) during both training and inference, named baseline (1/4sub), baseline (1/6sub), and baseline (1/8sub). The only difference between these baseline models and HydraFormer lies in the utilization of independent subsampling modules, where each module processes a specific subsampling rate. Apart from this, 
the encoder, decoder, and other structures, as well as training and inference parameters and settings, remain consistent with HydraFormer to ensure a fair comparison.

For a comprehensive evaluation of the experimental results, we employ two decoding approaches: CTC greedy search on the Conformer-based encoder's output and attention rescoring on the BiTransformer-based decoder's output. We then measure the recognition results using the word error rate (WER) between the predicted text and the target transcription as the evaluation metric. More details can be found in the GitHub repository.

\vspace{-3mm}
\subsection{Impact of positional encoding on HydraFormer}
\vspace{-1mm}
This experiment on AISHELL-1 aims to investigate whether incorporating positional encoding (rel\_pos) or not (no\_pos) has an impact on the performance of the HydraFormer model.

\setcounter{table}{0} 
\begin{table}[htbp]
\centering
\vspace{-1mm}
\caption{Positional encoding's impact on HydraFormer's performance using various subsampling rates on AISHELL-1.}
\vspace{-2mm}
\scalebox{0.89}{
\begin{tabular}{ccc}
\toprule[2pt]
Model                      & WER (CTC) & WER (Rescore) \\ \hline
HydraFormer rel\_pos (1/4sub) & 6.20      & 5.12          \\
HydraFormer no\_pos (1/4sub)  & 5.99      & 4.99          \\ \hline
HydraFormer rel\_pos (1/6sub) & 6.24      & 5.24          \\
HydraFormer no\_pos (1/6sub)  & 6.04      & 5.02          \\ \hline
HydraFormer rel\_pos (1/8sub) & 6.34      & 5.29          \\
HydraFormer no\_pos (1/8sub)  & 6.24      & 5.17        \\
\bottomrule[2pt]
\end{tabular}
}
\vspace{-3mm}
\end{table}

\begin{table*}[ht]
\caption{Comparison of HydraFormer and baseline models on LibriSpeech and AISHELL-1 datasets. WERs for CTC greedy search and attention rescore are represented as ``X/Y" in the table, where X is the CTC greedy search WER and Y is the attention rescore WER. ``1/nsub" indicates the model trained with a single subsampling rate of $n$. The train cost of an independent model is set to 1, consistent with the train cost of HydraFormer, while the combined train cost for three baseline models is 3.}
\vspace{-2mm}
\centering
\scalebox{0.798}{
\begin{tabular}{c|c|ccc|ccc|ccc}
\toprule[2pt]
{\multirow{3}{*}{Model / Params}} & \multirow{3}{*}{\begin{tabular}[c]{@{}c@{}}Train\\  Cost\end{tabular}} & \multicolumn{3}{c|}{Subsampling4}                          & \multicolumn{3}{c|}{Subsampling6}                          & \multicolumn{3}{c}{Subsampling8}                           \\
{}                       &                                                                         & \multicolumn{2}{c}{LibriSpeech} & \multirow{2}{*}{AISHELL} & \multicolumn{2}{c}{LibriSpeech} & \multirow{2}{*}{AISHELL} & \multicolumn{2}{c}{LibriSpeech} & \multirow{2}{*}{AISHELL} \\
{}                       &                                                                         & test-clean     & test-other     &                          & test-clean     & test-other     &                          & test-clean     & test-other     &                          \\ \hline
baseline 1/4sub / 48.3M     & \multirow{3}{*}{3}                                                      & 4.73/3.92      & 11.48/10.21    & 5.94/5.00                & \textemdash              & \textemdash              & \textemdash                        & \textemdash              & \textemdash              & \textemdash                        \\
baseline 1/6sub / 48.9M     &                                                                         & \textemdash              & \textemdash              & \textemdash                        & 4.87/3.91      & 11.83/10.48    & 6.03/5.22                & \textemdash              & \textemdash              & \textemdash                        \\
baseline 1/8sub / 48.3M    &                                                                         & \textemdash              & \textemdash              & \textemdash                        & \textemdash              & \textemdash              & \textemdash                        & 5.04/4.01      & 12.17/10.63    & 6.01/5.16                \\ \hline
{HydraFormer / 51.7M}            & 1                                                                       & 4.72/3.80      & 11.91/10.33    & 5.99/4.99                & 4.93/3.86      & 12.34/10.70    & 6.04/5.02                & 5.57/4.21      & 13.13/11.32    & 6.24/5.17      \\
\bottomrule[2pt]
\end{tabular}
}
\vspace{-4mm}
\end{table*}

As shown in Table 1, at subsampling rate 4, HydraFormer without positional encoding (no\_pos) outperforms its counterpart with positional encoding (rel\_pos), achieving WER of 5.99\% (CTC) and 4.99\% (Rescore) compared to 6.20\% (CTC) and 5.12\% (Rescore). This suggests that omitting positional encoding can improve performance.
For subsampling rates 6 and 8, similar trend can be observed. HydraFormer without positional encoding (no\_pos) consistently outperforms its counterpart with positional encoding (rel\_pos). This further supports the absence of positional encoding leads to superior performance in the context of HydraFormer.

Based on these findings, it is evident that not using positional encoding in HydraFormer results in better performance across different subsampling rates on AISHELL-1 dataset. This can be attributed to the fact that the encoder's temporal perception ability might be negatively affected by the use of positional encoding in conjunction with different subsampling rates in HydraFormer. By omitting positional encoding, the HydraFormer model's adaptability and performance across various application scenarios can be enhanced, further emphasizing the model's potential for optimization and improvements. Therefore, in all subsequent experiments, we do not use positional encoding in the HydraFormer model.

\vspace{-3mm}
\subsection{Main results on AISHELL-1 and LibriSpeech}
\vspace{-1mm}
To thoroughly evaluate the effectiveness of HydraFormer in handling multiple subsampling tasks within a single model while maintaining exceptional recognition performance, we conduct experiments on both English LibriSpeech dataset and Chinese Mandarin AISHELL-1 dataset, by comparing the results with three individual baseline models where each is trained separately on 4, 6, and 8 subsampling rate branches. The results are shown in Table 2.

For LibriSpeech dataset, HydraFormer attains competitive WER results across all subsampling rates, showcasing its adaptability without compromising performance. At subsampling rate 4, HydraFormer achieves WER of 4.72\% (CTC) and 3.80\% (Rescore), indicating the model can adapt to the subsampling rate of 4 without performance degradation. At subsampling rate 6, HydraFormer exhibits WER of 4.93\% (CTC) and 3.86\% (Rescore), closely matching the baseline model. Although the performance gap widens at subsampling rate 8, HydraFormer maintains reasonable WER of 5.57\% (CTC) and 4.21\% (Rescore), demonstrating it can still cater to higher subsampling rates with competitive performance.

For AISHELL-1 dataset, HydraFormer also demonstrates competitive performance across various subsampling rates, emphasizing its adaptability to different languages and datasets. At subsampling rate 4, it attains WER of 5.99\% (CTC) and 4.99\% (Rescore), which is marginally higher in CTC but lower in Rescore to the baseline model, highlighting the model's adaptability across languages. At subsampling rate 6, HydraFormer achieves WER of 6.04\% (CTC) and 5.02\% (Rescore), on par with the baseline model, further showcasing its adaptability across various subsampling rates. At subsampling rate 8, HydraFormer reaches WER of 6.24\% (CTC) and 5.17\% (Rescore), slightly higher than the baseline model, but still maintaining competitive performance, consistent with the observations in the LibriSpeech experiment.


It is crucial to emphasize that HydraFormer reduces the train cost to \textbf{1/3} compared to the baseline models 
adhering to 
only one subsampling rate. This adaptability is apparent on both LibriSpeech and AISHELL-1 datasets, indicating that HydraFormer can efficiently cater to the requirements of diverse contexts and languages without incurring additional costs associated with training multiple models for distinct scenarios. In conclusion, HydraFormer exhibits a remarkable capability to adapt to varying subsampling rates and languages while preserving exceptional recognition performance, all with a reduced training cost.

\vspace{-3mm}
\subsection{Experiments on recognition speed}
\vspace{-1mm}

To evaluate HydraFormer's performance in terms of recognition speed across varying subsampling rates, we use Real-Time Factor (RTF) as the metric in both streaming and non-streaming settings on the AISHELL-1 dataset for experiment.

\begin{table}[h]
\setcounter{table}{2} 
\vspace{-1mm}
\caption{Comparison of Real-Time Factor (RTF) for HydraFormer using various subsampling rates on AISHELL-1 dataset in streaming and non-streaming settings on an Intel Xeon Platinum 8255C CPU @ 2.50GHz with 1 thread.}
\vspace{-2mm}
\centering
\scalebox{0.9}{
\begin{tabular}{ccc}
\toprule[2pt]
      subsampling rate              & streaming & non-streaming \\ \hline
4 & 0.136113  & 0.114975      \\
6 & 0.127044  & 0.096963 \\ 
8 & 0.116731  & 0.087345     \\
\bottomrule[2pt]
\end{tabular}
}
\vspace{-2mm}
\end{table}

As shown in Table 3, HydraFormer shows an increasing efficiency as the subsampling rate increases, illustrated by a decreasing trend in RTF. In the streaming scenario, HydraFormer exhibits RTF of 0.136113 at subsampling rate 4, which decreases to 0.127044 and 0.116731 at subsampling rates 6 and 8, respectively. Similarly, in the non-streaming setting, the RTF decreases from 0.114975 at subsampling rate 4 to 
0.096963
and 0.087345 at subsampling rates 6 and 8.

\vspace{-3mm}
\subsection{Analysis of tradeoff between accuracy \& speed and real world application of HydraFormer}
\vspace{-1mm}

Referencing Tables 2 and 3, it is observed HydraFormer's recognition performance decreases (i.e. WER increases) and recognition speed increases (i.e. RTF decreases) with higher subsampling rates. This observation is consistent with the underlying rationale of HydraFormer, which highlights the importance of considering multiple criteria when evaluating speech recognition models, rather than solely relying on WER.  When other factors remain constant, a lower WER typically implies a higher RTF. However, different application scenarios have varying requirements: some prioritize recognition speed, while others emphasize recognition accuracy. Therefore, it is essential to design multiple models tailored to specific use cases, catering to the diverse needs of real-world situations. For example, consider a hybrid cloud-edge system. The cloud model is trained with a subsampling rate of 4, as accuracy is the paramount metric in the cloud scenario. Meanwhile, the edge model is trained with a subsampling rate of 8, with speed as the primary concern. The edge side is responsible for quickly generating results, tolerating certain errors, as the final outcomes will be replaced by the more accurate results from the cloud. Traditionally, this would require training two independent models, doubling the training cost. By employing HydraFormer, one can configure HydraSub as a combination of subsampling rates 4 and 8. In this setup, a single model can adapt to both cloud and edge scenarios, effectively reducing the training cost by 50\% compared to the conventional approach.

\vspace{-3mm}
\subsection{Experiments on different initialization strategies}
\vspace{-1mm}

To explore the impact of various initialization strategies on HydraFormer, we further conduct experiments on AISHELL-1 dataset. For a more intuitive presentation of the experimental results, we divide HydraFormer into two parts: HydraSub and Encoder+Decoder.

\begin{table}[h]
\centering
\vspace{-1mm}
\caption{Impact of various initialization strategies on HydraFormer.
Different branches of HydraSub can be separately initialized. 
``s'' denotes training from scratch, ``n'' represents transferring weights from ``baseline (1/nsub)''.}
\vspace{-2mm}
\scalebox{0.63}{
\begin{tabular}{c|cc|ccc}
\toprule[2pt]
\multirow{2}{*}{ID} & \multicolumn{2}{c|}{HydraFormer Initialization} & \multirow{2}{*}{\begin{tabular}[c]{@{}c@{}}Subsampling4\\ (CTC/Rescore)\end{tabular}} & \multirow{2}{*}{\begin{tabular}[c]{@{}c@{}}Subsampling6\\ (CTC/Rescore)\end{tabular}} & \multirow{2}{*}{\begin{tabular}[c]{@{}c@{}}Subsampling8\\ (CTC/Rescore)\end{tabular}} \\
                    & HydraSub    & Encoder+Decoder    &                                                                                       &                                                                                       &                                                                                       \\ \hline
\#1                 & s\_s\_s     & from scratch       & 5.99/4.99                                                                             & 6.04/5.02                                                                             & 6.24/5.17                                                                             \\ \hline
\#2                 & s\_s\_s     & baseline (1/4sub)          & 5.94/5.06                                                                             & 5.99/5.12                                                                             & 6.28/5.28                                                                             \\
\#3                 & 4\_s\_s     & baseline (1/4sub)         & 5.96/5.05                                                                             & 5.99/5.11                                                                             & 6.20/5.25                                                                             \\
\#4                 & 4\_s\_s     & from scratch       & 5.87/5.02                                                                             & 5.95/5.13                                                                             & 6.09/5.17                                                                             \\ \hline
\#5                 & 4\_6\_8     & baseline (1/4sub)          & 5.92/5.06                                                                             & 5.96/5.09                                                                             & 5.87/5.22                                                                             \\
\#6                 & 4\_6\_8     & from scratch       & 5.82/4.94                                                                             & 5.82/5.02                                                                             & 5.87/4.99               \\
\bottomrule[2pt]
\end{tabular}
}
\vspace{-3mm}
\end{table}

\textbf{Initialize from a single subsampling rate model (\#2 to \#4)}: Considering that a significant portion of current ASR models are trained using a single subsampling rate, we evaluate the effectiveness of initializing HydraFormer using a model trained solely on a fixed subsampling rate. As shown in Table 4, when initializing HydraFormer with baseline (1/4sub) for both HydraSub-4 and Encoder+Decoder (\#3), it can achieve competitive WER for all subsampling rates of 4, 6, and 8. A similar trend is observed between \#2 and \#4, suggesting that initializing HydraFormer with a single subsampling rate model is applicable and can maintain stable performance across various subsampling rates, highlighting its adaptability and effectiveness in diverse scenarios.

\textbf{Initialize from multiple subsampling rate models (\#5 \& \#6)}: We further investigate the advantages of initializing HydraFormer with weights transferred from multiple pretrained models, each corresponding to a specific subsampling rate. Our findings indicate that this strategy can enhance performance, as the model is able to leverage prior knowledge from each subsampling rate. 
When initializing with weights from baseline (1/4sub), baseline (1/6sub), and baseline (1/8sub) for their respective HydraSub branches, HydraFormer (\#6) achieves WER of 5.82\% (CTC) and 4.94\% (Rescore) for subsampling rate 4, outperforming 
the results 
when trained from scratch (\#1). 
Furthermore, improved performances for subsampling rates of 6 and 8 are also observed,
illustrating HydraFormer's adaptability and effectiveness across various subsampling rates. 
When initializing the encoder and decoder with baseline (1/4sub) (\#5), we note that although there is a slight decline in performance compared to random initialization (\#6), the results remain relatively close to those achieved by the entire model trained from scratch (\#1). This highlights the potential of using multiple subsampling rate models for initialization in diverse scenarios.

In conclusion, experiment results underscore the stability and robustness of HydraFormer, as it consistently maintains good performance across various initialization strategies. HydraFormer can be conveniently fine-tuned from a pretrained single subsampling rate model, maintaining stable performance and reducing training costs. Furthermore, transferring weights from pretrained ASR models with different subsampling rates to corresponding HydraSub branches does not cause interference. Instead, HydraFormer leverages the prior knowledge from these models to enhance its performance and adaptability across different subsampling rates. This demonstrates the strong adaptability, universality, and scalability of HydraFormer, making it suitable for a wide range of application scenarios.

\vspace{-3mm}
\subsection{Comparison of HydraFormer's parameters with single subsampling rate ASR models}
\vspace{-1mm}

To provide a more intuitive understanding of the relationship between the HydraFormer model and the models trained using individual subsampling rates, we visualize the encoder parameters of the models trained on the AISHELL-1 dataset, including baseline (1/4sub), baseline (1/6sub), baseline (1/8sub), and HydraFormer. The encoder comprises 12 Conformer blocks, and for an in-depth analysis of the parameter distribution, we focus on two representative layers within the Conformer blocks: a linear layer and a convolutional layer. Specifically, we select portions of the parameters in the linear and convolutional layers that share the same position across all 12 blocks and use t-SNE\cite{van2008tsne} to reduce the representation to 2 dimensions.
\begin{figure}[h]
\vspace{-2mm}
    \centering
    \begin{tabular}{@{\extracolsep{\fill}}c@{}c@{\extracolsep{\fill}}}
            \includegraphics[width=0.5\linewidth]{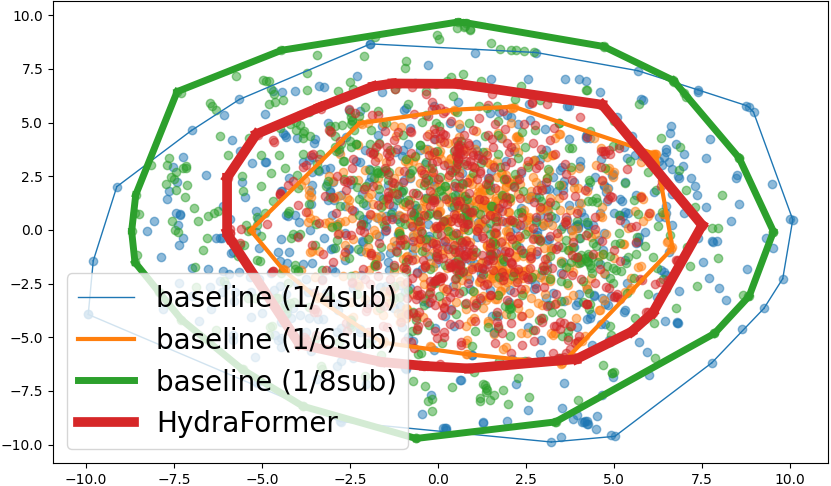} &
            \includegraphics[width=0.5\linewidth]{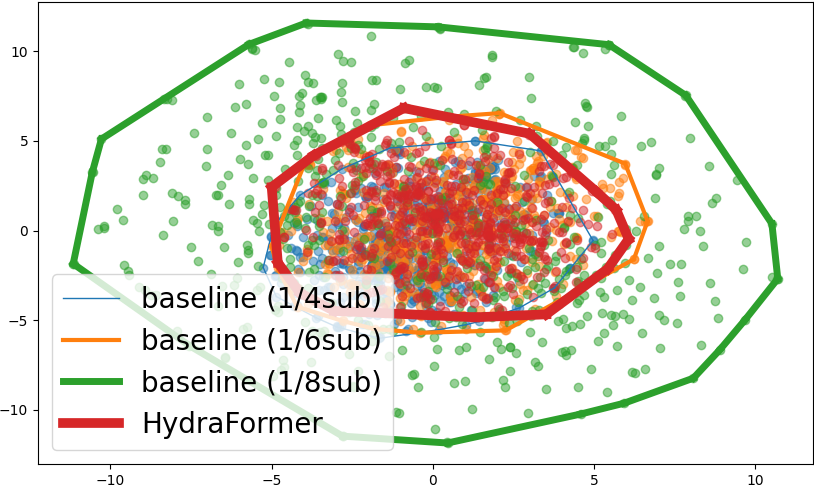}\\
            (a) Convolutional layer & (b) Linear layer\\
    \end{tabular}
    \vspace{-2mm}
    \caption{Parameter distribution visualization of HydraFormer and baseline models}
    \label{fig:image_with_table}
    \vspace{-5mm}
 \end{figure}
 
As depicted in Fig. 3, we observe that the parameter distribution of HydraFormer consistently lies near the intersection of the parameters of the three baseline models trained with subsampling rates of 4, 6, and 8, i.e., baseline (1/4sub), baseline (1/6sub), and baseline (1/8sub). This observation holds true for both convolutional and linear layer parameter plots. Despite variations in plot structures stemming from the distinct roles of the convolutional and linear layers, the overall trend remains consistent across both layers.

The visualization highlights that HydraFormer effectively captures the essential characteristics of each baseline model, demonstrating its adaptability in managing various subsampling rates within a single, unified model. The consistent trend observed across different layers and Conformer blocks further underscores the model's ability to learn and adjust its internal representations throughout the entire encoder, enhancing its effectiveness in diverse application scenarios.

\vspace{-1mm}
\section{Conclusion}
\vspace{-2mm}
To tackle the issue of handling multiple different subsampling rates within a single model and reduce deployment and training costs, we introduce HydraFormer. This versatile model effectively replaces several individual models, significantly lowering the expenses associated with training and deployment while maintaining high recognition performance. The experiments conducted on AISHELL-1 and LibriSpeech datasets demonstrate that HydraFormer achieves nearly equivalent performance at different subsampling rates compared to models trained only at specific subsampling rates. Furthermore, HydraFormer exhibits stability across various model initialization strategies.

\vspace{-1mm}
\section{Acknowledgment}
\vspace{-2mm}
This work is supported by Shenzhen Science and Technology Program (WDZC20200818121348001, JCYJ202208181010\\14030) and the Major Key Project of PCL (PCL2022D01, PCL2023AS7-1).

\bibliographystyle{IEEEbib}
\ninept
\bibliography{icme2023template}

\end{document}